\documentclass[ALICE,manyauthors]{cernphprep}
\usepackage[comma,square,numbers,sort&compress]{natbib}

\usepackage{multirow}
\usepackage{lineno}
\usepackage{xspace}
\usepackage{hyperref}
\usepackage{color}
\usepackage[T1]{fontenc}
\usepackage{orcidlink}

\begin{document}
%

\newcommand{\pp}           {pp\xspace}
\newcommand{\ppbar}        {\mbox{$\mathrm {p\overline{p}}$}\xspace}
\newcommand{\XeXe}         {\mbox{Xe--Xe}\xspace}
\newcommand{\PbPb}         {\mbox{Pb--Pb}\xspace}
\newcommand{\pA}           {\mbox{pA}\xspace}
\newcommand{\pPb}          {\mbox{p--Pb}\xspace}
\newcommand{\AuAu}         {\mbox{Au--Au}\xspace}
\newcommand{\dAu}          {\mbox{d--Au}\xspace}

\newcommand{\s}            {\ensuremath{\sqrt{s}}\xspace}
\newcommand{\snn}          {\ensuremath{\sqrt{s_{\mathrm{NN}}}}\xspace}
\newcommand{\pt}           {\ensuremath{p_{\rm T}}\xspace}
\newcommand{\meanpt}       {$\langle p_{\mathrm{T}}\rangle$\xspace}
\newcommand{\ycms}         {\ensuremath{y_{\rm CMS}}\xspace}
\newcommand{\ylab}         {\ensuremath{y_{\rm lab}}\xspace}
\newcommand{\etarange}[1]  {\mbox{$\left | \eta \right |~<~#1$}}
\newcommand{\yrange}[1]    {\mbox{$\left | y \right |~<~#1$}}
\newcommand{\dndy}         {\ensuremath{\mathrm{d}N_\mathrm{ch}/\mathrm{d}y}\xspace}
\newcommand{\dndeta}       {\ensuremath{\mathrm{d}N_\mathrm{ch}/\mathrm{d}\eta}\xspace}
\newcommand{\avdndeta}     {\ensuremath{\langle\dndeta\rangle}\xspace}
\newcommand{\dNdy}         {\ensuremath{\mathrm{d}N_\mathrm{ch}/\mathrm{d}y}\xspace}
\newcommand{\Npart}        {\ensuremath{N_\mathrm{part}}\xspace}
\newcommand{\Ncoll}        {\ensuremath{N_\mathrm{coll}}\xspace}
\newcommand{\dEdx}         {\ensuremath{\textrm{d}E/\textrm{d}x}\xspace}
\newcommand{\RpPb}         {\ensuremath{R_{\rm pPb}}\xspace}

\newcommand{\nineH}        {$\sqrt{s}~=~0.9$~Te\kern-.1emV\xspace}
\newcommand{\seven}        {$\sqrt{s}~=~7$~Te\kern-.1emV\xspace}
\newcommand{\twoH}         {$\sqrt{s}~=~0.2$~Te\kern-.1emV\xspace}
\newcommand{\twosevensix}  {$\sqrt{s}~=~2.76$~Te\kern-.1emV\xspace}
\newcommand{\five}         {$\sqrt{s}~=~5.02$~Te\kern-.1emV\xspace}
\newcommand{\twosevensixnn}{$\sqrt{s_{\mathrm{NN}}}~=~2.76$~Te\kern-.1emV\xspace}
\newcommand{\fivenn}       {$\sqrt{s_{\mathrm{NN}}}~=~5.02$~Te\kern-.1emV\xspace}
\newcommand{\LT}           {L{\'e}vy-Tsallis\xspace}
\newcommand{\GeVc}         {Ge\kern-.1emV/$c$\xspace}
\newcommand{\MeVc}         {Me\kern-.1emV/$c$\xspace}
\newcommand{\TeV}          {Te\kern-.1emV\xspace}
\newcommand{\GeV}          {Ge\kern-.1emV\xspace}
\newcommand{\MeV}          {Me\kern-.1emV\xspace}
\newcommand{\GeVmass}      {Ge\kern-.2emV/$c^2$\xspace}
\newcommand{\MeVmass}      {Me\kern-.2emV/$c^2$\xspace}
\newcommand{\lumi}         {\ensuremath{\mathcal{L}}\xspace}

\newcommand{\ITS}          {\rm{ITS}\xspace}
\newcommand{\TOF}          {\rm{TOF}\xspace}
\newcommand{\ZDC}          {\rm{ZDC}\xspace}
\newcommand{\ZDCs}         {\rm{ZDCs}\xspace}
\newcommand{\ZNA}          {\rm{ZNA}\xspace}
\newcommand{\ZNC}          {\rm{ZNC}\xspace}
\newcommand{\SPD}          {\rm{SPD}\xspace}
\newcommand{\SDD}          {\rm{SDD}\xspace}
\newcommand{\SSD}          {\rm{SSD}\xspace}
\newcommand{\TPC}          {\rm{TPC}\xspace}
\newcommand{\TRD}          {\rm{TRD}\xspace}
\newcommand{\VZERO}        {\rm{V0}\xspace}
\newcommand{\VZEROA}       {\rm{V0A}\xspace}
\newcommand{\VZEROC}       {\rm{V0C}\xspace}
\newcommand{\Vdecay} 	   {\ensuremath{V^{0}}\xspace}

\newcommand{\ee}           {\ensuremath{e^{+}e^{-}}} 
\newcommand{\pip}          {\ensuremath{\pi^{+}}\xspace}
\newcommand{\pim}          {\ensuremath{\pi^{-}}\xspace}
\newcommand{\kap}          {\ensuremath{\rm{K}^{+}}\xspace}
\newcommand{\kam}          {\ensuremath{\rm{K}^{-}}\xspace}
\newcommand{\pbar}         {\ensuremath{\rm\overline{p}}\xspace}
\newcommand{\kzero}        {\ensuremath{{\rm K}^{0}_{\rm{S}}}\xspace}
\newcommand{\lmb}          {\ensuremath{\Lambda}\xspace}
\newcommand{\almb}         {\ensuremath{\overline{\Lambda}}\xspace}
\newcommand{\Om}           {\ensuremath{\Omega^-}\xspace}
\newcommand{\Mo}           {\ensuremath{\overline{\Omega}^+}\xspace}
\newcommand{\X}            {\ensuremath{\Xi^-}\xspace}
\newcommand{\Ix}           {\ensuremath{\overline{\Xi}^+}\xspace}
\newcommand{\Xis}          {\ensuremath{\Xi^{\pm}}\xspace}
\newcommand{\Oms}          {\ensuremath{\Omega^{\pm}}\xspace}
\newcommand{\degree}       {\ensuremath{^{\rm o}}\xspace}

\newcommand{\dedx}{d$E$/d$x$}
\newcommand{\dndydpt}{${\rm d}^2N/({\rm d}y {\rm d}p_{\rm T})$}
\newcommand{\pbpb}{Pb--Pb}
\newcommand{\sig}{$\Sigma^{+}$}
\newcommand{\asig}{$\overline{\Sigma}^{-}$}

\newcommand{\sigplusasig}{$\Sigma^{+}+$\hspace{0.6mm}$\overline{\Sigma}^{-}$}

\begin{titlepage}
\PHyear{2025}       
\PHnumber{181}      
\PHdate{11 Aug}  

\title{$\Sigma^{+}$ production in pp collisions at $\mathbf{\sqrt{\textit s}=13}$~TeV}
\ShortTitle{$\Sigma^{+}$ production in pp collisions at $\sqrt{s}=13$~TeV}   

\Collaboration{ALICE Collaboration\thanks{See Appendix~\ref{app:collab} for the list of collaboration members}}
\ShortAuthor{ALICE Collaboration} 

\begin{abstract}
The measurement of \sig\ production in pp collisions at $\sqrt{s}=13$~TeV is presented. The measurement is performed at midrapidity in both minimum-bias and high-multiplicity pp collisions at $\s=13$~TeV. 
The \sig\ is reconstructed via its weak-decay topology in the decay channel $\Sigma^{+} \rightarrow \rm{p} + \pi^{0}$ with $\pi^{0} \rightarrow \gamma + \gamma$. 
In a novel approach, the neutral pion is reconstructed by combining photons that convert in the detector material with photons measured in the calorimeters. 
The transverse-momentum ($p_{\rm T}$) distributions of the \sig\ and its rapidity densities ${\rm d}N$/${\rm d}y$ in both event classes are reported. The $p_{\rm T}$ spectrum in minimum-bias collisions is compared to QCD-inspired event generators. The ratio of \sig\ to previously measured $\Lambda$ baryons is in good agreement with calculations from the Statistical Hadronization Model. The high efficiency and purity of the novel reconstruction method for \sig\ presented here will enable future studies of the interaction of \sig\ with protons in the context of femtoscopic measurements, which could be crucial for understanding the equation of state of neutron stars.
\end{abstract}
\end{titlepage}

\setcounter{page}{2} 


\section{Introduction} 

Over the last five decades, the production of many different particles has been measured with high precision in hadronic and heavy-ion reactions. Even short-lived particles and rare probes could be studied. However, the members of the $S=-1$ isospin triplet in the $J^P=\frac{1}{2}^+$ SU(3) baryon octet, the $\Sigma$ baryons, are scarce. 
This is because most experiments are optimized for the detection of decays into charged particles and none of the ground-state $\Sigma$ baryons has a purely charged decay channel. In any case, either a neutron or photon(s) occur in the final state. The photons stemming from the decays typically have low energy and are therefore difficult to detect in the multi-particle final state of a hadronic collision. 

The study of baryons with strangeness makes an important contribution to the understanding of the strangeness production mechanisms in pp collisions and provides a reference for QCD-inspired Monte Carlo (MC) event generators, which generally tend to underestimate the yield of strange hadrons. 
In pp collisions at LHC energies a significant tension was observed between the measured transverse momentum ($p_{\rm T}$) distributions of $\Lambda$ baryons and the expectations from MC generators\,\cite{ATLAS:2011xhu,ALICE:2020jsh}. In previous measurements of excited states of $\Sigma$ baryons, comparisons were made with different MC generators and large deviations in both the shape and yield of the $p_{\rm T}$ spectra were found\,\cite{ALICE:2022wpn,ALICE:2014zxz,ALICE:2017pgw,ALICE:2022zuc}. In this respect, the measurement of \sig\ is an important cross-check and an input to constrain the models, e.g., PYTHIA\,\cite{Skands:2010ak,Skands:2014pea} and EPOS\,\cite{Pierog:2013ria,Werner:2023zvo}. 

Strange baryon production has also been studied in $e^{+}e^{-}$ collisions at LEP. In particular, the DELPHI\,\cite{DELPHI:1995dso,DELPHI:2000oqt}, OPAL\,\cite{OPAL:1996dbo,OPAL:2009hyp}, and L3\,\cite{L3:2000wwp} Collaborations measured the production of ground-state $\Sigma$ baryons. Only recently, ALICE measured the anti-particles of the charged $\Sigma$ baryons for the first time in pp and p--Pb collisions at $\sqrt{s_\textrm{NN}}=5.02$~TeV\,\cite{ALICE:2025yxi}. They compared it with available event generators and found a deviation similar to the one observed for other strange baryons and a characteristic ordering of the strange baryon yield with mass and strangeness content.

The \sig\ baryon was measured by the LHCb Collaboration in pp collisions at $\sqrt{s}=7$, 8, and 13~TeV to determine the branching fraction of the rare semi-leptonic decay into a proton and a dimuon\,\cite{LHCb:2017rdd, LHCb:2025evf}. They found the branching fraction to be compatible with the standard model expectation but did not report the production cross section of the \sig\ baryons.

The mass dependence of particle yields can be explained in the framework
of statistical hadronization models (SHM). In heavy-ion collisions, the SHM description is based on a grand-canonical ensemble and has three main parameters: the chemical freeze-out temperature $T_{\rm ch}$, the baryochemical potential $\mu_{B}$ and the volume $V$ of the fireball. The baryochemical potential $\mu_{B}$ is a measure of the net-baryon content of the system and is expected to be close to zero at the LHC, i.e., baryons and antibaryons are produced in equal amounts. The SHM provides a very good description of particle yields in heavy-ion collisions over a wide range of collision energies and particle species\,\cite{Andronic:2010qu, Andronic:2017pug, ALICE:2017jmf}. A canonical formulation of the SHM\,\cite{Becattini:1995if,Becattini:2008tx} is used in smaller systems such as in $e^{+}e^{-}$ or pp collisions. The canonical formulation requires an additional parameter $R_C$, which corresponds to the radius within which the exact conservation of quantum numbers, e.g., strangeness, is enforced. In more recent implementations of the SHM, this parameter is often expressed in terms of a correlation volume $V_C$, which can be calculated from $R_C$. In some implementations, a non-equilibrium parameter $\gamma_S$\,\cite{Koch:1986ud,Rafelski:1991rh} is additionally used, that helps to describe the undersaturation of strangeness in small systems\,\cite{Vovchenko:2019kes}.
Remarkably, in contrast to the microscopic MC generators, the SHM\,\cite{Becattini:1995if,Becattini:2008tx} works very well for the strange baryon production yields at LEP.

Furthermore, the measurement of the \sig\ production is important as an input for the measurement of inclusive charged-particle spectra. Since measured \sig\ spectra are not available and the production of strange baryons is not well reproduced in MC generators, their contribution is currently inferred from measured \pt spectra of $\Lambda$\,\cite{ALICE:2018vuu,ALICE:2018hza}.

In this paper, results on the production of the \sig\ baryon (and its antiparticle \asig) in pp collisions at \mbox{$\s=13$~TeV} are reported for the first time. This measurement is made possible thanks to a novel reconstruction method in the $\Sigma^{+} \rightarrow \rm{p} + \pi^{0}$ $(\pi^{0} \rightarrow \gamma + \gamma)$ decay channel (branching ratio ($51.57\pm 0.30$)~\%\,\cite{ParticleDataGroup:2020ssz}), where the two decay photons are reconstructed via a conversion $\gamma \rightarrow \rm{e}^{+} + \rm{e}^{-}$ and a direct measurement in the calorimeter, respectively. Unless otherwise stated, all considerations regarding \sig\, apply analogously to \asig, using the respective charge-conjugate daughter particles.

\section{Data analysis}
\label{sec:data_ana}

A detailed description of the ALICE detectors and their performance can be found in\,\cite{Aamodt:2008zz,ALICE:2014sbx}. Within the scope of this paper, only a brief description of the used detectors will be given. Those are the Inner Tracking System (ITS), the Time Projection Chamber (TPC), the Time-Of-Flight detector (TOF), the electromagnetic calorimeters (PHOS, ECal, DCal), as well as the V0 detectors. 

The ITS is a six-layer cylindrical silicon detector used for tracking, vertexing, and triggering. The layers are located at radii between 3.9 and 43 cm around the beam axis. The inner two layers of the ITS are also used for triggering. The TPC is a cylindrical gaseous detector located around the ITS with an inner radius of around 85 cm and an outer radius of 250 cm. The TPC is the main tracking detector and also contributes to the determination of the primary vertex. Furthermore, the TPC is used for particle identification (PID) through the measurement of the specific energy loss in the detector gas. The ITS and the TPC are located inside a 0.5 T strong solenoidal magnetic field and cover a pseudorapidity range of $|\eta| < 0.9$ in the full azimuth. The TOF detector wraps around the Transition Radiation Detector (TRD), which envelops the TPC. The TOF is used as a supplementary tool for PID through the measurement of the particle velocity. The PHOton Spectrometer (PHOS) and the Electromagnetic Calorimeter (EMCal) are calorimeters forming the outermost part of the central barrel detector. They enhance the capabilities to measure jets and electromagnetic probes such as photons. The EMCal consists of two sub-detectors, the ECal and the Di-jet Calorimeter (DCal). ECal and DCal are lead/polystyrene sampling calorimeters covering 110\degree and 60\degree in azimuthal angle and are placed on opposite sides in azimuth. Both have a common pseudorapidity coverage of $|\eta| < 0.7$. The PHOS detector has a higher granularity and is made of lead-tungstate crystals. It is placed between the DCal modules. The PHOS covers a pseudorapidity range of $|\eta| < 0.12$ and 70\degree in azimuth. In contrast to ECal and DCal, PHOS is placed behind a hole in the TRD and TOF detectors to minimize the material budget.

The V0 detectors consist of two plastic scintillator arrays located at forward $(2.8 < \eta < 5.1)$ and backward $(-3.7 < \eta < -1.7)$ pseudorapidities and are used for event triggering. The minimum bias trigger used in this analysis requires at least one hit in both V0 detectors. The number of triggered events corresponds to the V0 visible cross section, which can be related to the total inelastic pp cross section. The amplitude of the signal in the V0 detectors is proportional to the multiplicity; if it exceeds a certain threshold, the event is triggered as a high-multiplicity one.

The data analyzed in this paper was recorded between 2016 and 2018 in the LHC pp run at \s = 13 TeV. The analysis was performed using a high-multiplicity (HM) and a minimum-bias (MB) triggered sample. The recorded integrated luminosity amounts to 0.059~pb$^{-1}$ using the minimum-bias trigger and 14.1~pb$^{-1}$ using the high-multiplicity triggers\,\cite{ALICE:2023udb}. After the event selection, about $1.8\cdot 10^9$ MB triggered events and $1.3\cdot 10^9$ HM triggered events were available for the analysis.

Tuning of the particle selection criteria and the correction for the reconstruction efficiency is carried out using an MC simulation. Due to the low reconstruction efficiency of \sig\ and \asig, in particular at low \pt, a dedicated simulation with enhanced \sigplusasig yield is used, such that the corresponding statistical uncertainty is not a dominant contribution in the final analysis. To this end, underlying events generated with PYTHIA 8 Monash 2013\,\cite{Skands:2014pea} are enriched with 10 \sig\ and 10 \asig following a L$\acute{\textrm{e}}$vy-Tsallis\,\cite{Tsallis:1987eu, STAR:2008med} transverse momentum distribution. Additionally, one \sig\ and one \asig are injected with a uniform \pt\ distribution from 1 to 10~GeV/$\it{c}$. The parameters of the L$\acute{\textrm{e}}$vy-Tsallis\,\cite{Tsallis:1987eu, STAR:2008med} transverse momentum distribution are obtained by fitting a preliminary \sig\ spectrum that was corrected using a general-purpose simulation. The described injection scheme leads to an approximate uniform \pt\ distribution of the reconstructed spectrum. The simulated particles are propagated through the detectors using GEANT4\,\cite{Agostinelli:2002hh}.

The \sig\ is reconstructed in the decay mode $\Sigma^{+} \rightarrow \rm{p} + \pi^{0}$, with $\pi^{0} \rightarrow \gamma + \gamma$. 
The protons are selected by their specific energy loss in the TPC gas (d$E$/d$x$) which must be within 3$\sigma$ of its expected value. For protons with transverse momenta greater than 0.9 GeV/$c$, an additional 5$\sigma$ selection is made on the time-of-flight signal to enhance the purity. To avoid compromising efficiency, the selection is only applied to tracks which have a hit in the TOF detector. Only tracks within a pseudorapidity range of $|\eta| < 0.9$ are considered. The distance-of-closest approach (DCA) in the $xy$ plane (perpendicular to the beam direction) is required to be greater than 10 µm to suppress primary particles, and less than 2 cm to suppress protons knocked out of the detector material.

The neutral pion is reconstructed by combining a photon that converts in the detector material and a photon measured in the calorimeters. Photons can convert into an electron-positron pair in the presence of a nucleus within the detector material. The reconstruction of photons from such conversions is called photon conversion method (PCM). The advantage of the PCM is that it gives access to topological information which allows for the high purities of the particle sample. The combination with the calorimeters allows to make use of their significantly higher detection efficiency. 

The PCM photons are reconstructed from $e^{+}$-$e^{-}$ pairs during the tracking and selected by their distinct V-shaped topology (V$^0$). 
Only a minimal selection is applied to the electron tracks, which must have a d$E$/d$x$ within 5$\sigma$ of the expected value, and at least 30 (out of 159) space points in the TPC. The opening angle of the V$^0$, i.e., the angle between the electron tracks at the point of the conversion, must be less than 0.3 rad. The magnetic field in the detector points in the $z$ direction (along the beam direction), so the angle between a track and the magnetic field is given by the $\theta$ coordinate. The difference in this coordinate between the electron tracks must be less than 0.1~rad. The invariant mass of the photon candidate must be less than 0.06 GeV/$c^2$. In addition, selections are made in the Armenteros-Podolanski\,\cite{Arm_Pod} plane to separate photons from $\Lambda$ baryons and $\rm K^{0}_{\rm S}$ mesons, which also decay in a V$^0$ topology. In the Armenteros-Podolanski plane, the longitudinal momentum asymmetry $\alpha=(p_l^+-p_l^-)$/$(p_l^++p_l^-)$ of the daughter tracks is related to the transverse momentum $q_\text{T}$ of the daughters with respect to the flight direction of the V$^0$. PCM photons are selected if $|\alpha| < 0.9$ and $q_{\text{T}} < 0.03 ~{\rm GeV}/\it{c}$.

For the second photon the EMCal and PHOS calorimeters are used. 
In contrast to the photons reconstructed by the PCM, the number of applicable selections for the calorimeter photons is limited, resulting in lower purity. Since the photons from the \sig\ decays have low energy, they deposit most of it in a single calorimeter cell, which hinders the application of a selection on the cluster shape. 
Track matching is performed and clusters are discarded if at least one charged-particle track is within 10~cm from the cluster position in the calorimeter plane, effectively suppressing clusters caused by charged particles.

The decay vertex of the \sig\ is reconstructed using the Kalman Filter (KF) package\,\cite{KFParticle}. For this purpose, only the proton track and the PCM photon are used, since the momentum components of the calorimeter photon are unknown. After the computation of the secondary vertex, the flight path of the calorimeter photon is constructed as the connecting line between the vertex and the position of the respective calorimeter cluster. From the reconstructed secondary vertex, the pointing angle ($PA$) of the \sig\ is computed as

\begin{align}
PA = \arccos\left(\frac{\vec{r}\cdot\vec{p}}{|\vec{r}|\cdot|\vec{p}|}\right),
\label{PA_eq}
\end{align}

where $\vec{r}$ is the decay vertex position with respect to the primary vertex and $\vec{p}$ is the momentum vector.

The reconstructed \sig\ are identified by their considerably long lifetime of about 80~ps ($c\tau\approx2.4 \text{ cm}$), which allows strong selections on the decay topology. 
The distance between the reconstructed decay vertex of the \sig\ and the primary vertex must be greater than 1 cm. The strong momentum dependence of the $PA$ resolution gives rise to a momentum-dependent selection. The pointing angle must be less than 0.012~rad for $p_{\rm T} > 2 \,{\rm \,GeV}/\it{c}$ and less than 0.024~rad for $p_{\rm T} < 2 \,{\rm \,GeV}/\it{c}$.

The moderate energy resolution of the calorimeters leads to a significant broadening of the resulting invariant-mass distribution of the \sig. To avoid this, an energy correction is applied based on the reconstructed topology. First, the invariant mass of the $\pi^0$ is calculated using the measured calorimeter energy and the momentum of the PCM photon. If the reconstructed mass is within the range 0.1--0.16~GeV/$\it{c}^2$, the corrected energy of the calorimeter photon ($E_{\text{Calo}}$) is calculated as

\begin{align}
E_{\text{Calo}} = \frac{m_{\pi^0}^2}{2\cdot|\vec{p}_{\text{PCM}}|\cdot(1-\cos{\alpha})},
\label{Ecorr_eq}
\end{align}

where $m_{\pi^0}$ is the nominal $\pi^0$ mass, $\vec{p}_{\text{PCM}}$ is the momentum of the PCM photon and $\alpha$ is the angle between the two photons. 

The combinatorial background of the reconstructed \sig\ is described by using the event-mixing technique. For this, the tuples of protons and conversion photons are mixed with calorimeter clusters from different events and fed into the same reconstruction procedure as in the same event, and the same selections are applied. To account for efficiency and acceptance effects, only events that have a similar primary-vertex position in the $z$-coordinate ($\pm$1 cm) as well as a similar multiplicity ($\pm$4 tracks in $|\eta|<0.8$) are paired. Up to 10 events are mixed to improve the statistics. The mixed-event distribution is fitted to the same-event distribution in the sideband region by scaling it with a momentum-dependent parameter. 

\begin{figure}[!ht]
\centering
\includegraphics[width=0.49\textwidth]{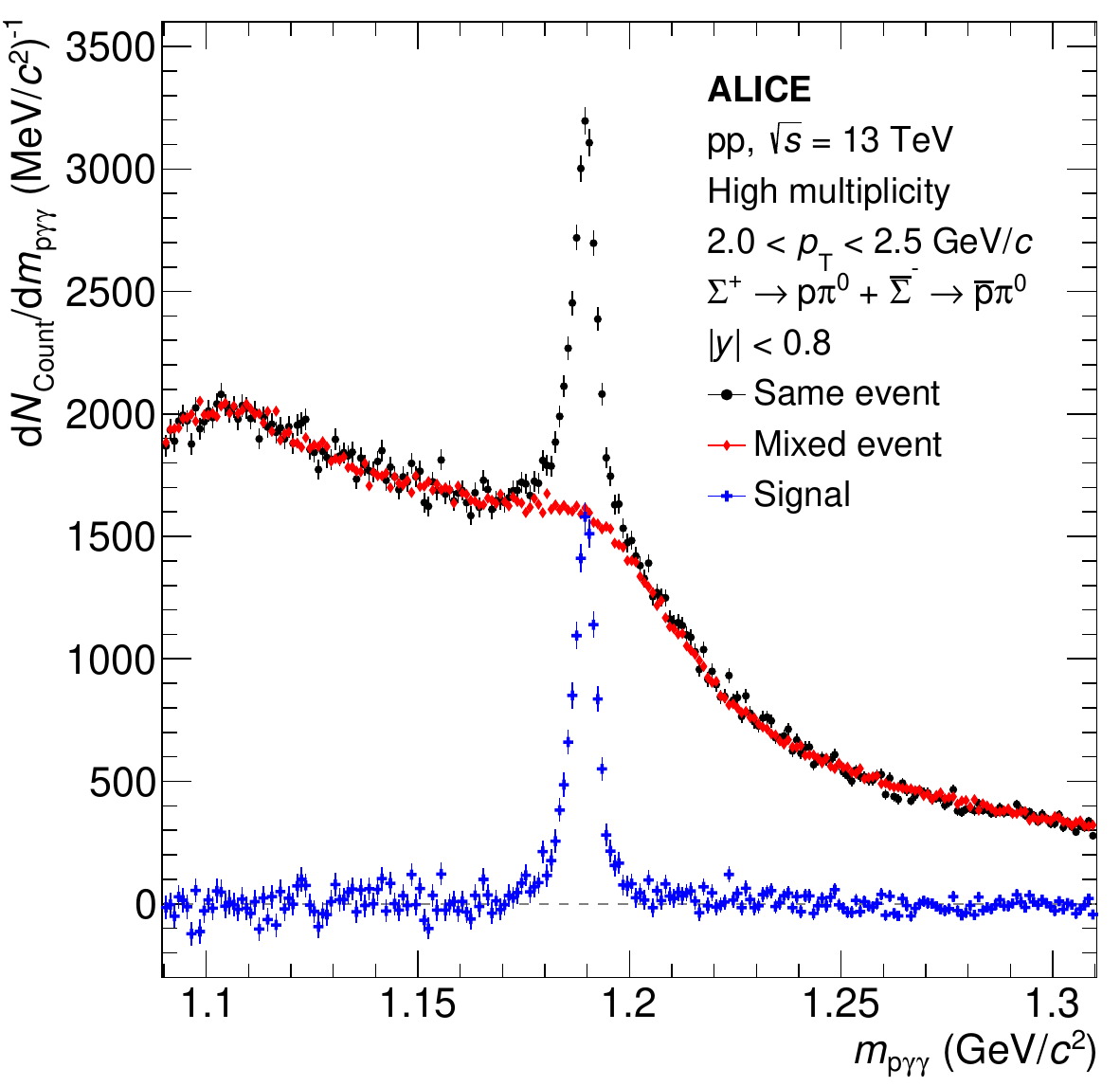}
\includegraphics[width=0.49\textwidth]{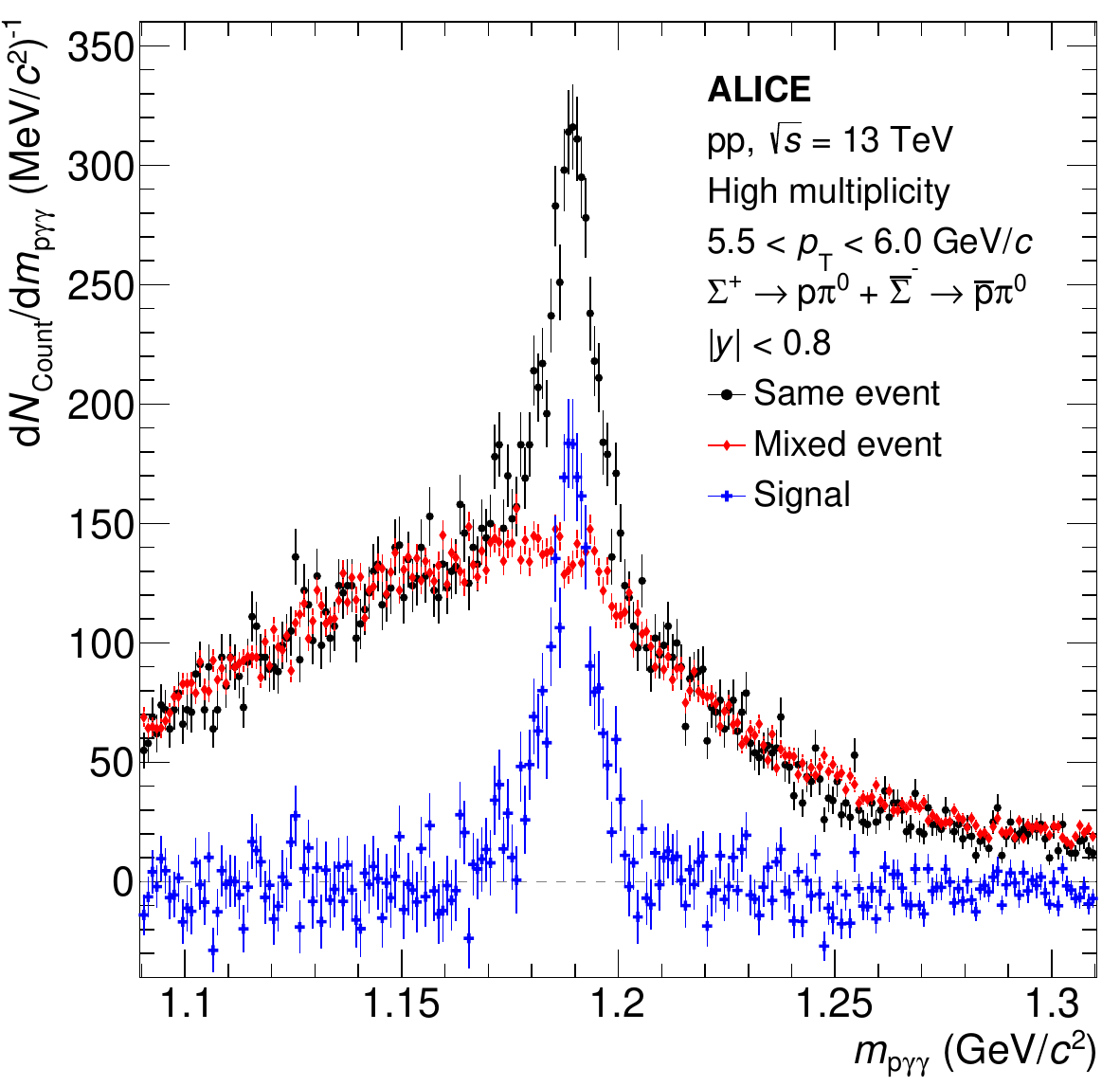}
\includegraphics[width=0.49\textwidth]{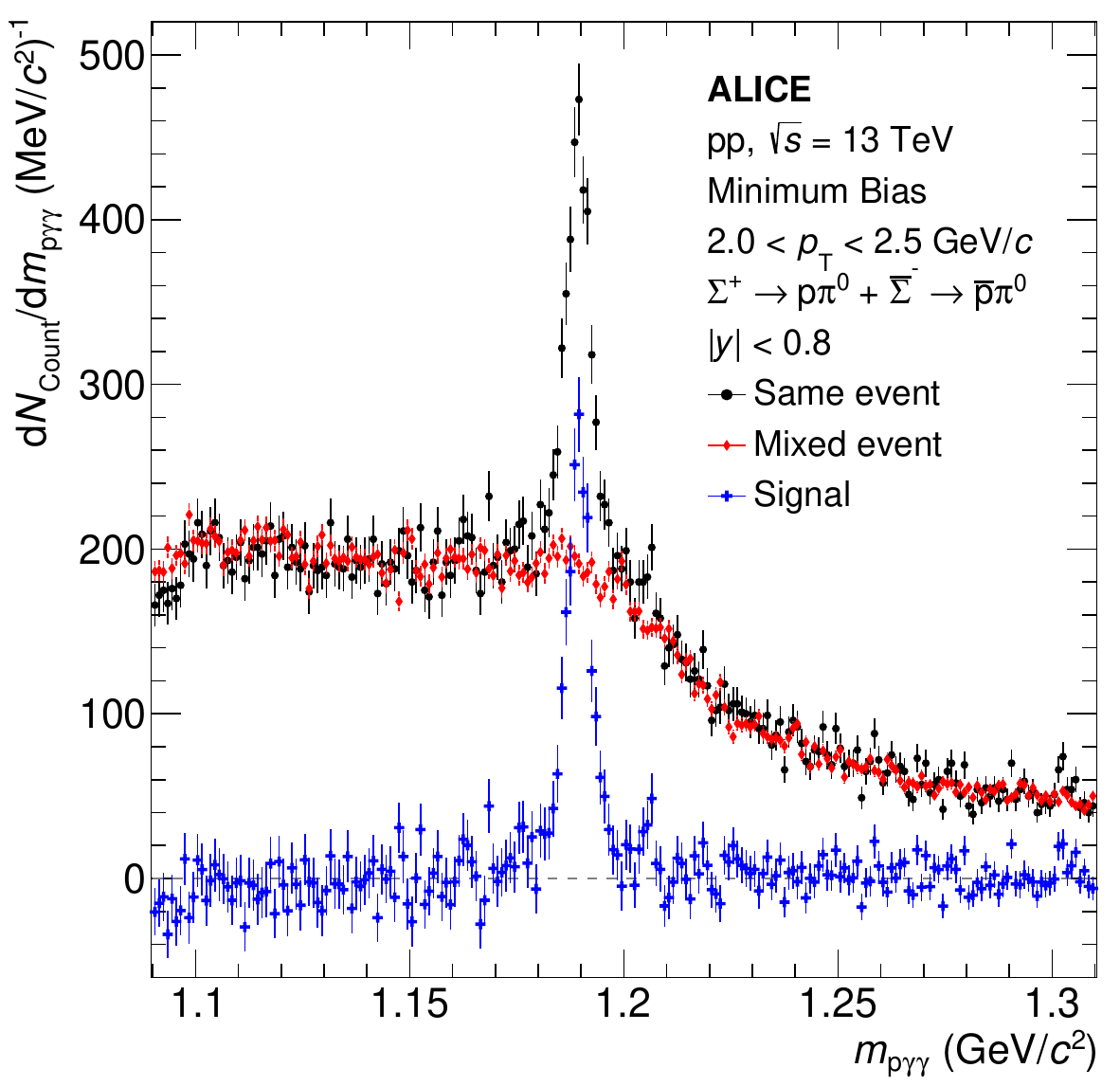}
\includegraphics[width=0.49\textwidth]{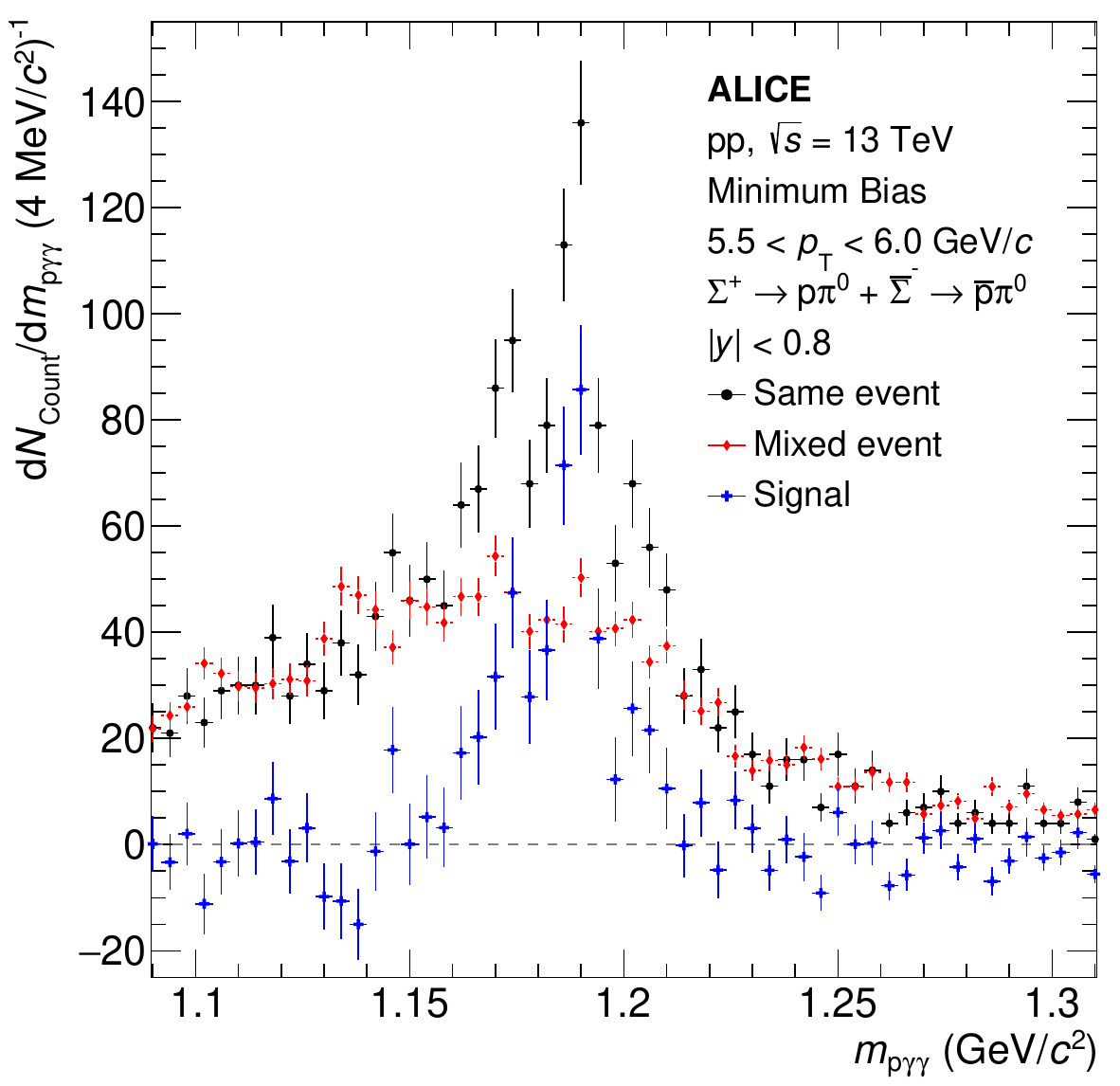}
\caption{Invariant-mass distributions of \sigplusasig candidates in high-multiplicity (upper panels) and minimum-bias (lower panels) pp collisions at $\sqrt{s}=13$~TeV in 2.0 $< p_{\rm T} <$ 2.5~$\rm GeV/\it{c}$ (left panels) and in 5.5 $< p_{\rm T}\ <$ 6.0~$\rm GeV/\it{c}$ (right panels) in same events (black symbols), mixed events (red symbols) and signal (blue symbols). The error bars of the data points correspond to the statistical uncertainties.}
\label{invmass_se_me}
\end{figure}

Figure~\ref{invmass_se_me} shows the same-event (black points), mixed-event (red points), and subtracted (blue points) p$\gamma\gamma$ invariant-mass distributions. The background of the same-event distribution is well described by the mixed events. The signal is obtained by subtracting the mixed-event distribution from the same-event one, as seen in Fig.~\ref{invmass_se_me}. The nominal mass of the \sig\ of 1.189 \GeVmass\,\cite{ParticleDataGroup:2020ssz} is well reproduced and a good agreement of the peak width with the reconstructed MC signal is found. The number of counts in each $p_{\rm T}$ interval is determined by bin counting of the signal distributions within a 3$\sigma$ interval around the nominal \sig\ mass.

\begin{figure}[!htb]
\centering
\includegraphics[width=0.7\textwidth]{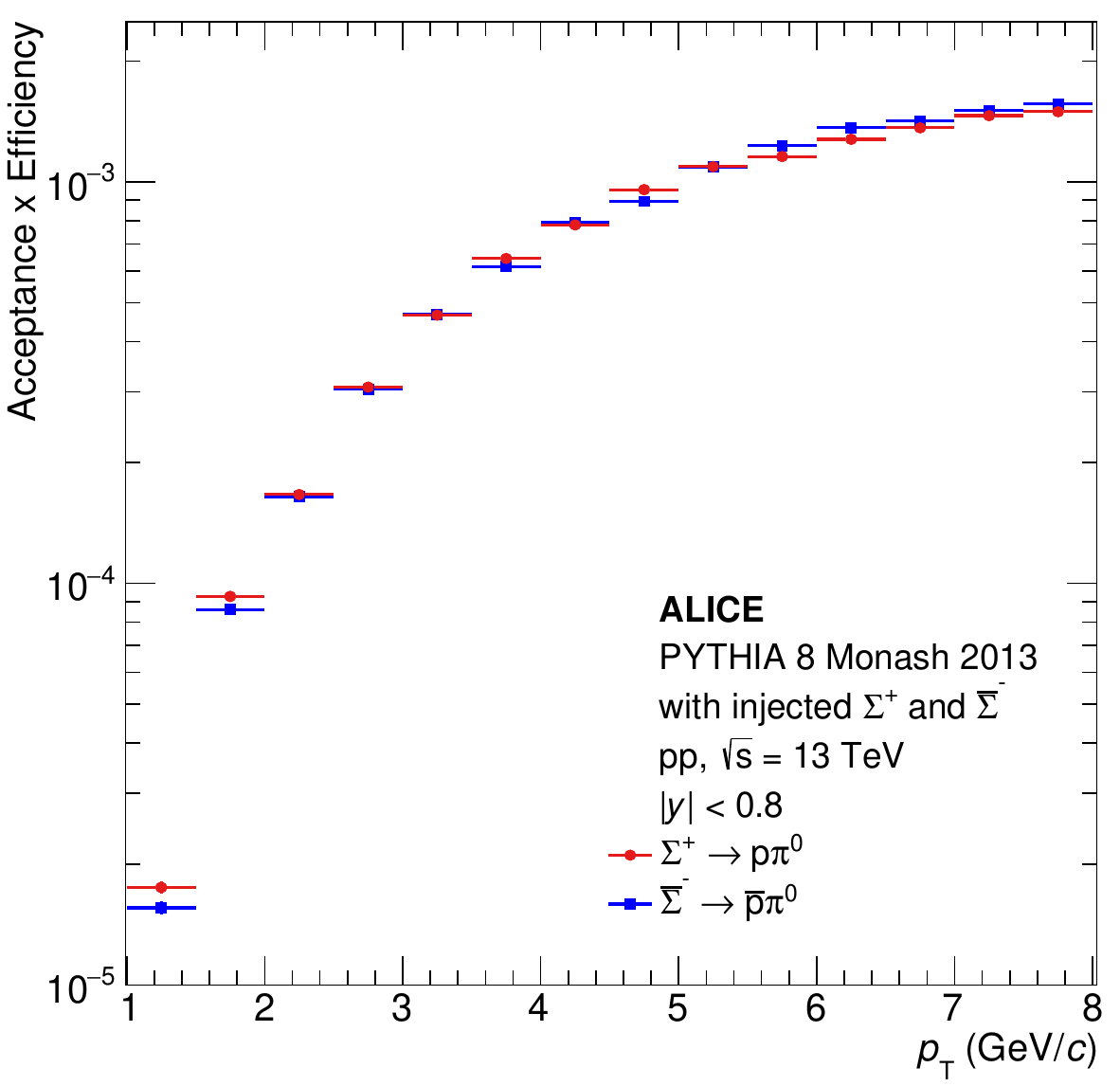}
\caption{Acceptance $\times$ efficiency as a function of the transverse momentum of \sig\ and \asig\ after applying the selection criteria described in the text. The results for \sig\ (red symbols) and \asig\ (blue symbols) are shown. The error bars of the data points correspond to the statistical uncertainties.}
\label{reco_eff}
\end{figure}

The obtained raw spectra are corrected for the acceptance and reconstruction efficiency which are determined from the MC simulations. The minimum-bias spectrum is additionally corrected for the efficiency of the V0 detectors such that it corresponds to the total inelastic pp cross section. The acceptance $\times$ efficiency is shown in Fig.~\ref{reco_eff}. It is strongly $p_{\rm T}$ dependent and decreases towards lower $p_{\rm T}$ due to the low photon momenta. The multiplicity dependence of the reconstruction efficiency for different pp collision classes is negligible, so the same simulations can be used for the MB-triggered and the HM-triggered data sets.

The measured $p_{\rm T}$ spectra are subject to systematic uncertainties. In order to quantify these uncertainties, variations of the selections are made. In order to take into account potential correlations among the selection variables, the variations are not performed individually but in groups. These groups are track, PID, kinematic, and topological selections. Each selection is varied within half a unit of standard deviation from the default value, if applicable, or 10\% variation is used otherwise. For each selection group, different sets of selections are randomly chosen within the defined ranges. The selections are thereby chosen from a continuous uniform distribution, i.e., each setting is equally likely. In total 1000 selection sets are analyzed. For each variation, the raw spectrum and the corresponding efficiencies are evaluated and the deviation of the efficiency corrected spectrum from the default setting is calculated. 
The deviations are Gaussian distributed and are fitted to determine the mean and the width. Therefore, this method is insensitive to outliers. The distributions are not necessarily centered at zero, which gives rise to asymmetrical error bars. In addition to the systematic uncertainties which arise from the selection variations, other uncertainties are added which are assumed to be independent of the momentum. These uncertainties originate from the conversion probability, i.e., the material budget, the ITS-TPC matching efficiency and mixed-event (ME) normalization, where the given uncertainties are estimated based on previous analyses\,\cite{ALICE:2020nkc}. The cluster finding efficiency gives rise to an additional uncertainty. The selection variation uncertainties exhibit a slight momentum dependence. In the first \pt\ interval, the kinematic and topological selection uncertainties increase significantly, which is related to a particular sensitivity of the efficiency on these selections at low \pt\ and the limited statistics in this interval. The systematic uncertainties, summarized in Tab.~\ref{tab:syst_table}, are evaluated in the high-multiplicity sample due to the better statistical precision, but are very similar in the minimum-bias sample. The total systematical uncertainty is taken as the quadratic sum of the systematics of all contributions. 

\newcommand\ChangeRTv[1]{{\vrule width #1}}
\begin{table}[!htb]
\centering
\caption{Systematical and statistical uncertainties of the corrected yields of \sigplusasig given in \%.}
\begin{tabular}{|ccc|c|c|c|c|c|}
\hline
$p_{\textrm T}$ (GeV/$\textit{c}$)         &\ChangeRTv{1.5pt}& 1.0-2.0 & 2.0-3.0 & 3.0-4.0 & 4.0-5.0 & 5.0-6.0 & 6.0-7.0 \\
\hline\hline
Topological             &\ChangeRTv{1.5pt}& 2.3 & 0.8 & 0.6 & 0.9 & 1.2 & 1.6 \\
Kinematic               &\ChangeRTv{1.5pt}& 4.2 & 1.8 & 2.6 & 2.0 & 3.1 & 3.6 \\
Track selection         &\ChangeRTv{1.5pt}& 0.2 & 0.2 & 0.2 & 0.1 & 0.2 & 0.2 \\
Particle identification &\ChangeRTv{1.5pt}& 1.1 & 0.3 & 0.3 & 0.3 & 0.3 & 0.8 \\
Material budget         &\ChangeRTv{1.5pt}& 4.5 & 4.5 & 4.5 & 4.5 & 4.5 & 4.5 \\
ITS-TPC matching        &\ChangeRTv{1.5pt}& 4.0 & 4.0 & 4.0 & 4.0 & 4.0 & 4.0 \\
Cluster efficiency      &\ChangeRTv{1.5pt}& 3.0 & 3.0 & 3.0 & 3.0 & 3.0 & 3.0 \\
ME normalisation        &\ChangeRTv{1.5pt}& 2.0 & 2.0 & 2.0 & 2.0 & 2.0 & 2.0 \\
Total systematic        &\ChangeRTv{1.5pt}&  8.7 & 7.3 & 7.5 & 7.4 & 7.8 & 8.0  \\ \hline\hline
Statistical (HM)        &\ChangeRTv{1.5pt}&  1.5 & 0.9 & 1.0 & 1.3 & 2.0 & 3.0  \\
Statistical (MB)        &\ChangeRTv{1.5pt}&  3.2 & 2.2 & 2.8 & 4.0 & 5.7 & 9.2  \\ \hline
\end{tabular}
\label{tab:syst_table}
\end{table}

\section{Results}

\begin{figure}[!htb]
\centering
\includegraphics[width=0.75\textwidth]{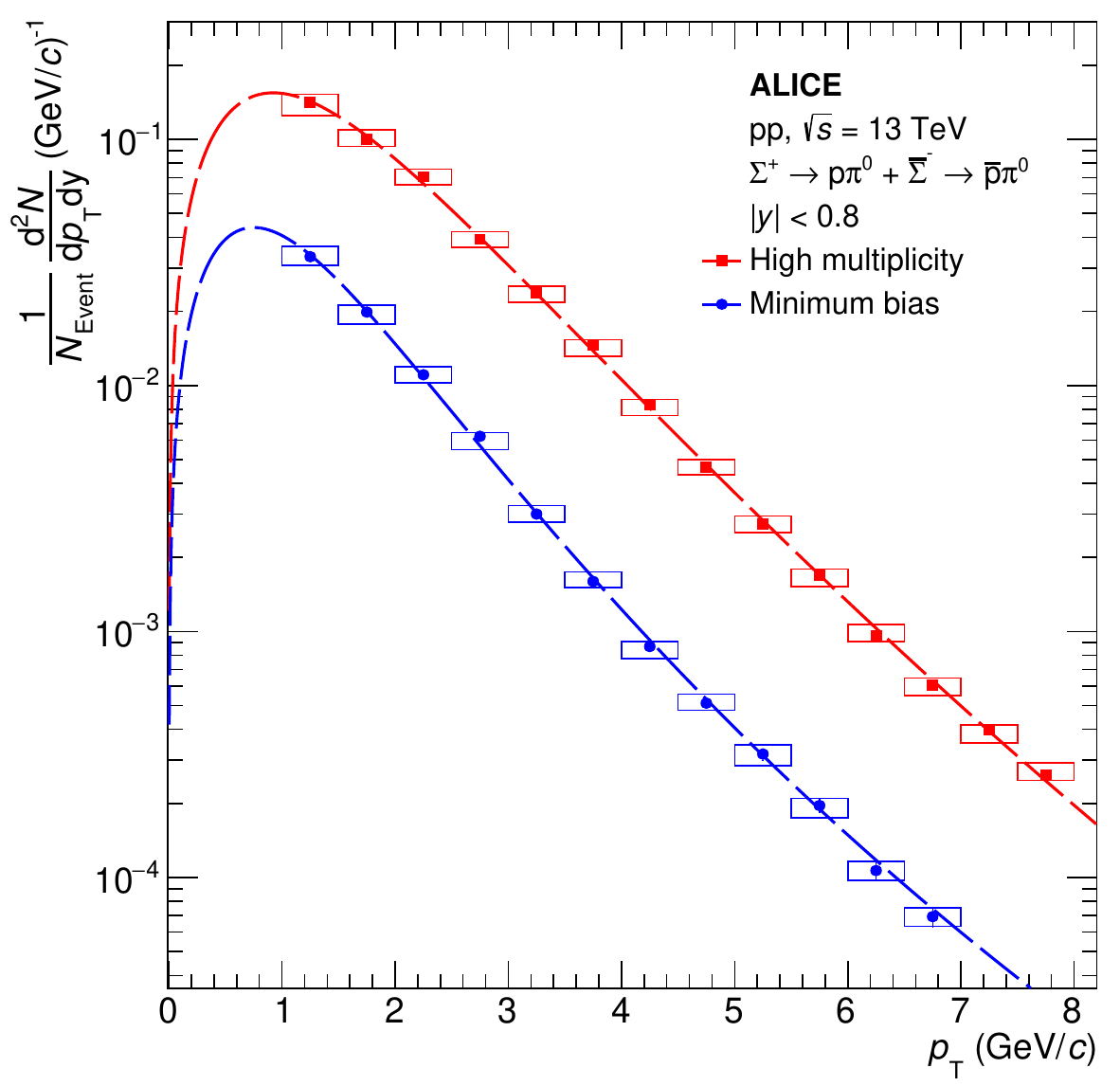}
\caption{Normalized differential yields of \sigplusasig as a function of $p_{\rm T}$ in high-multiplicity (red symbols) and minimum-bias (blue symbols) pp events at $\sqrt{s}=13$~TeV together with the corresponding L$\acute{\textrm{e}}$vy-Tsallis fits (dashed lines). The spectra are normalized to the number of events ($N_{\textrm{Event}}$) given in Section~\ref{sec:data_ana}. The statistical uncertainties of the data points are shown as bars and the systematic ones as boxes.}
\label{pt_spectra_withfits}
\end{figure}

The normalized differential yields of \sigplusasig as a function of $p_{\rm T}$ in high-multiplicity and minimum-bias pp collisions at $\sqrt{s}=13$~TeV are shown in Fig.~\ref{pt_spectra_withfits}. The different $p_{\rm T}$ range is due to the lower statistics in the MB sample. The spectral shape in the measured $p_{\rm T}$ intervals is well described by fits 
with L$\acute{\textrm{e}}$vy-Tsallis functions\,\cite{Tsallis:1987eu, STAR:2008med}, which are also shown in Fig.~\ref{pt_spectra_withfits}.

\begin{figure}[!htb]
\centering
\includegraphics[width=0.75\textwidth]{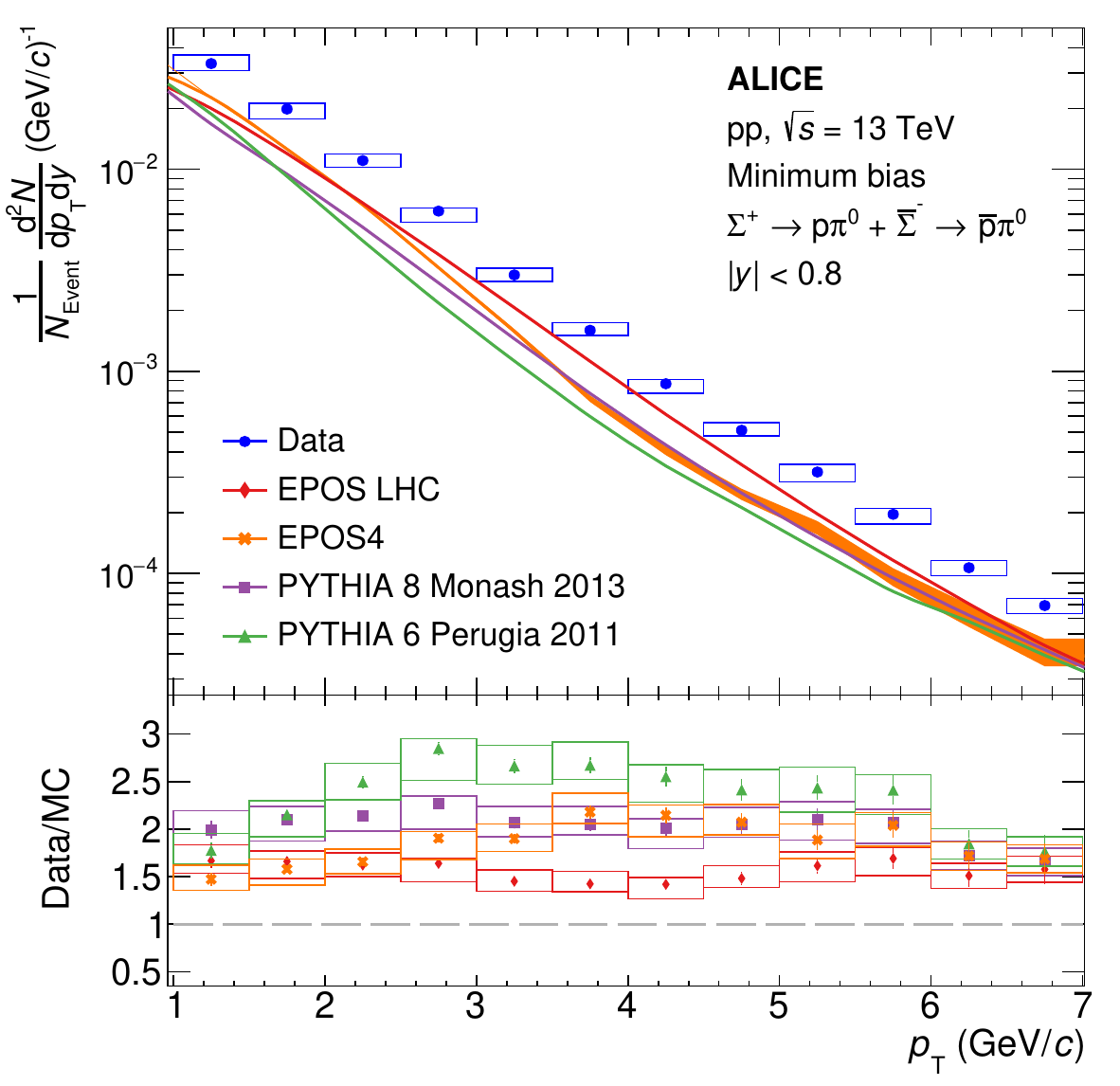}
\caption{The normalized differential yield of \sigplusasig as a function of $p_{\rm T}$ in minimum-bias pp collisions at $\sqrt{s}=13$~TeV (blue symbols) compared with PYTHIA 6 (green line), PYTHIA 8 (purple line), EPOS LHC (red line) and EPOS4 (orange line) generators. The lower panel shows the ratio between the data points and the MC predictions. The statistical uncertainties of the data points are shown as bars and the systematic ones as boxes. The statistical uncertainties of the model predictions are given by the width of the bands.}
\label{generator_comparison}
\end{figure}

In Fig.~\ref{generator_comparison}, the $p_{\rm T}$ spectrum of \sigplusasig in minimum-bias pp collisions is compared to spectra 
generated by the models PYTHIA 8~Monash 2013\,\cite{Skands:2014pea}, PYTHIA 6~Perugia 2011\,\cite{Skands:2010ak}, EPOS~LHC\,\cite{Pierog:2013ria}, as well as to the recently publicly released EPOS version 4 (EPOS4)\,\cite{Werner:2023zvo}.
PYTHIA 8 can reproduce the spectral shape with an accuracy of better than 20\% but underestimates the yield by a factor of about two, mainly due to an overall underestimation of strangeness production in the simulation. PYTHIA 6 is closer to the data than PYTHIA 8 in the first measured $p_{\rm T}$ interval but cannot describe the $p_{\rm T}$ shape resulting in the largest discrepancy with the data among the generators considered. EPOS~LHC describes the spectral shape similarly well as PYTHIA 8 while also being closest to the data in terms of yield. Additionally, the mean multiplicity of the events is closest to data in EPOS~LHC. EPOS4 shows a very similar performance to EPOS~LHC at low $p_{\rm T}$, but approaches PYTHIA 8 at intermediate $p_{\rm T}$. 

The difference between the PYTHIA and EPOS models might be related to their specific implementations. The EPOS generators feature a core-corona model\,\cite{Werner:2007bf, Werner:2023jps}, where the corona is characterized by string fragmentation, similar to the Lund model\,\cite{Andersson:1983ia} used by PYTHIA. This contribution dominates the high-\pt\ particle production, which could explain the increasing agreement among the considered models towards higher momentum. At lower \pt, the QGP-like core part of the EPOS model is influenced by coalescence, which could potentially increase the yields of \sig\ baryons compared to PYTHIA, improving the agreement with the data. 

The rapidity density ${\rm d}N$/${\rm d}y$ of \sigplusasig is determined by integrating the spectra shown in Fig.~\ref{pt_spectra_withfits} over $p_{\rm T}$. Since the data points do not extend to $p_{\rm T}$ = 0, the spectra must be extrapolated using a fit function. For this purpose, various fit functions are tested. The models with only two free parameters, namely Boltzmann, $m_{\rm T}$-exponential, $p_{\rm T}$-exponential and Fermi-Dirac fail to describe the data, showing deviations at low and high $p_{\rm T}$. The more sophisticated L$\acute{\textrm{e}}$vy-Tsallis\,\cite{Tsallis:1987eu, STAR:2008med} and Boltzmann-Gibbs\,\cite{Schnedermann:1993ws} functions both describe the data well. In agreement with the expectation for pp collisions, the L$\acute{\textrm{e}}$vy-Tsallis function describes the data best over the full considered $p_{\rm T}$ range. Hence, this function is chosen to extrapolate the spectra. In the extrapolation region, the functions are integrated numerically. To determine the statistical and systematic uncertainties, the spectra are randomly varied 1000 times from their default value according to a Gaussian distribution, where the standard deviation corresponds to the one of the data points. Each data point is varied individually. All such spectra are fitted resulting in a series of curves. The yield is calculated for all variations. Finally, the yields and the corresponding uncertainties of the measured and extrapolated \pt\ region are added to obtain the integrated yields. The contribution of the extrapolated yield to the total one is around 33\% for high-multiplicity events and 44\% for minimum-bias ones. This difference is due to the softer spectrum at lower multiplicities. The results for ${\rm d}N$/${\rm d}y$ in the two event classes are given in Tab.~\ref{table_results} together with the mean charged-particle multiplicities $\langle dN_{ch}/d{\eta} \rangle$ in $|\eta|<0.5$, which are $30.8$ in the high-multiplicity and $6.9$ in the minimum-bias event sample. 

\renewcommand{\arraystretch}{0.7}

\begin{table}[hbt!]
\centering
\caption{Rapidity density ${\rm d}N$/${\rm d}y$ of \sigplusasig in high-multiplicity and minimum-bias pp collisions at $\sqrt{s}=13$~TeV.}
\begin{tabular}{|c|c|c|}
\hline
& \multirow{2}{*}{$\langle dN_{ch}/d\eta \rangle$\,\cite{ALICE:2020swj} }& \multirow{2}{*}{$\frac{{\rm d}N}{{\rm d}y}$} \\
&&\\
\hline
\multirow{2}{*}{High-multiplicity pp} & \multirow{2}{*}{$30.8\pm0.4$} & \multirow{2}{*}{$0.313^{+0.007}_{-0.007}(\rm{stat})^{+0.027}_{-0.032}(\rm{syst})$}\\
&&\\
\multirow{2}{*}{Minimum-bias pp} & \multirow{2}{*}{$6.9\pm0.1$} & \multirow{2}{*}{$0.071^{+0.003}_{-0.003}(\rm{stat})^{+0.008}_{-0.008}(\rm{syst})$}\\
&&\\
\hline
\end{tabular}
\label{table_results}
\end{table}

\begin{figure}[!htb]
\centering
\includegraphics[width=0.75\textwidth]{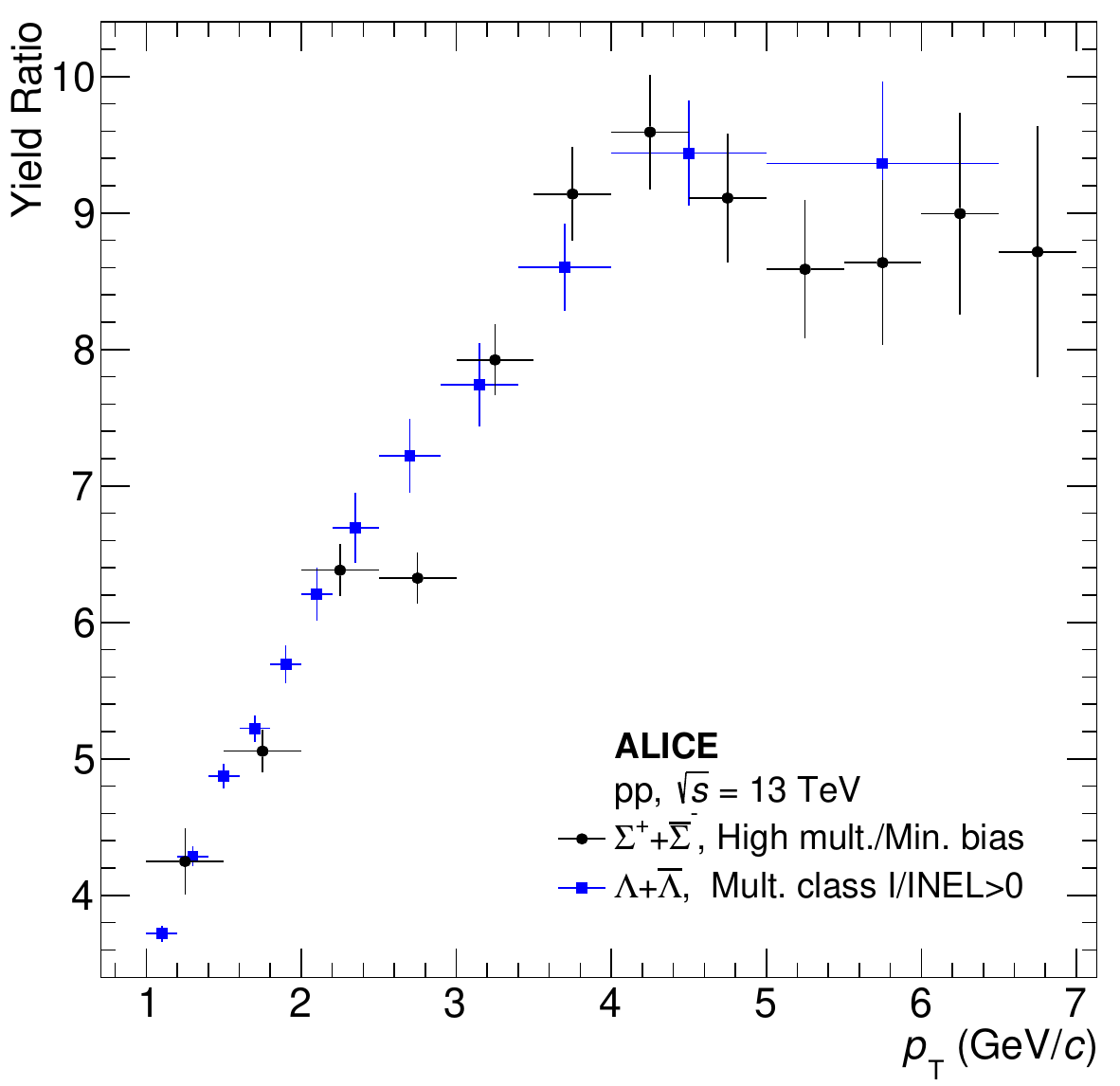}
\caption{Ratio of the \pt\ spectra of \sigplusasig in minimum-bias and high-multiplicity pp collisions. For comparison, the same ratio is shown for $\Lambda+\overline\Lambda$ baryons, taken from a previous analysis\,\cite{ALICE:2019avo}. One must note the slightly different multiplicities.}
\label{HM_MB_ratio}
\end{figure}

The ratio of the \pt\ spectra at low and high multiplicities is shown in Fig.~\ref{HM_MB_ratio}. An almost linear increase of the ratio with multiplicity is observed up to around $\pt =4$~GeV/$\it{c}$, followed by a saturation. For comparison, the same ratio is shown for $\Lambda$ baryons, taken from a previous analysis\,\cite{ALICE:2019avo}. Note that the definitions of the multiplicity classes are slightly different. A very good agreement of the ratios is apparent, which can be attributed to the similar mass of the considered baryons. 

The production yields of \sigplusasig are compared to those of $\Lambda+\overline\Lambda$ which have been measured in previous analyses\,\cite{ALICE:2019avo}. The data for $\Lambda+\overline\Lambda$ are extra- or interpolated to the mean charged-particle multiplicities of the event classes used in this analysis, assuming a linear dependence of the yield on the charged-particle multiplicity.

The ratios of the production yields of \sigplusasig and $\Lambda+\overline\Lambda$ baryons in both multiplicity classes are shown in Tab.~\ref{table_ratios}.
The observed values in the minimum-bias and high-multiplicity event classes are in good agreement within their uncertainties. This implies that the yield of $\Sigma$ baryons scales with multiplicity, like the yield of $\Lambda$ baryons, which in turn means that the same strangeness enhancement at higher multiplicity observed for the $\Lambda$\,\cite{ALICE:2016fzo} is also present for the $\Sigma$. 
A canonical SHM calculation\,\cite{Vovchenko:2019kes} gives a ratio of 0.285 for high-multiplicity and 0.271 for minimum-bias pp collisions, while a grand-canonical calculation\,\cite{Vovchenko:2019kes} gives 0.273. The data are in good agreement with both the canonical and the grand-canonical calculations.

\begin{table}[hbt!]
\centering
\caption{Yield ratios ($\Sigma^++\hspace{0.6mm}\overline{\Sigma}^-$)/($\Lambda+\overline\Lambda$) in high-multiplicity and minimum-bias pp collisions at $\sqrt{s}=13$~TeV and canonical calculations with SHM\,\cite{Vovchenko:2019kes}.}
\begin{tabular}
{|c|c|c|}
\hline
 \multirow{2}{*}{\raisebox{+1ex}{$\frac{\Sigma^++\hspace{0.6mm}\overline{\Sigma}^-}{\Lambda+\overline\Lambda}$}} & \multirow{2}{*}{ALICE measurements} & \multirow{2}{*}{Canonical SHM\,\cite{Vovchenko:2019kes}} \\
 &&\\
 \hline
\multirow{2}{*}{High-multiplicity pp} &\multirow{2}{*}{$0.275^{+0.006}_{-0.006}(\textrm{stat})^{+0.030}_{-0.034}(\textrm{syst})$}&\multirow{2}{*}{0.285}\\

&  &\\
\multirow{2}{*}{Minimum-bias pp}&  \multirow{2}{*}{$0.272^{+0.010}_{-0.010}(\textrm{stat})^{+0.035}_{-0.035}(\textrm{syst})$} & \multirow{2}{*}{0.271}\\
&&\\
\hline
\end{tabular}
\label{table_ratios}
\end{table}

\section{Summary and Conclusion}
This letter presents the measurement of the \sig\ and \asig production in pp collisions at $\s=13$~TeV at the LHC. Thanks to major innovations in the analysis technique, a large $\Sigma^{+}$ sample is reconstructed from the decay into a proton and a $\pi^0$, with the neutral pion being reconstructed from two photons. One of the photons is reconstructed with the photon conversion method and the other one is measured with signals from the electromagnetic calorimeters. The photon conversion method allows one to reconstruct the secondary vertex, which helps to improve the purity of the particle selection, while the calorimeters significantly improve the reconstruction efficiency.

The differential yields as a function of $p_{\rm T}$ are reported as well as the rapidity densities in minimum-bias and high-multiplicity pp collisions.
The experimental results are compared to Monte Carlo event generators. EPOS LHC is favoured by the data over PYTHIA 6 Perugia 2011, and PYTHIA 8 Monash 2013 tunes, as well as over EPOS4. The integrated yield ratios of $\Sigma$ to $\Lambda$ are in agreement with SHM calculations and suggest that the evolution of the $\Sigma$ production with multiplicity is compatible with that of $\Lambda$.
These observations are consistent with results from LEP, where the SHM approach was also used successfully~\cite{OPAL:1996dbo,L3:2000wwp,DELPHI:1995dso}, but the Monte Carlo generators could not reproduce the data.

The newly measured spectra of the $\Sigma$ baryon can be used to constrain the event generators further and can serve as an important input for the correction of inclusive charged-particle production measurements. 

The new analysis technique presented in this article provides significantly improved efficiency and purity of \sig\ baryon reconstruction compared to previous approaches. This opens up new possibilities for conducting studies based on clean $\Sigma$ measurements. Measuring the interaction of \sig\ with protons using femtoscopic measurements can provide important information about the equation of state of nuclear matter at high densities, as expected, for instance, in the inner core of neutron stars, where the strangeness degree of freedom might become relevant. $\Sigma$ baryons could also make an important contribution to baryon-strangeness correlations, which are related to studies of chemical freeze-out in heavy-ion collisions in the context of lattice QCD calculations\,\cite{Bellwied:2019pxh}.


\newenvironment{acknowledgement}{\relax}{\relax}
\begin{acknowledgement}
\section*{Acknowledgements}

The ALICE Collaboration would like to thank all its engineers and technicians for their invaluable contributions to the construction of the experiment and the CERN accelerator teams for the outstanding performance of the LHC complex.
The ALICE Collaboration gratefully acknowledges the resources and support provided by all Grid centres and the Worldwide LHC Computing Grid (WLCG) collaboration.
The ALICE Collaboration acknowledges the following funding agencies for their support in building and running the ALICE detector:
A. I. Alikhanyan National Science Laboratory (Yerevan Physics Institute) Foundation (ANSL), State Committee of Science and World Federation of Scientists (WFS), Armenia;
Austrian Academy of Sciences, Austrian Science Fund (FWF): [M 2467-N36] and Nationalstiftung f\"{u}r Forschung, Technologie und Entwicklung, Austria;
Ministry of Communications and High Technologies, National Nuclear Research Center, Azerbaijan;
Rede Nacional de Física de Altas Energias (Renafae), Financiadora de Estudos e Projetos (Finep), Funda\c{c}\~{a}o de Amparo \`{a} Pesquisa do Estado de S\~{a}o Paulo (FAPESP) and The Sao Paulo Research Foundation  (FAPESP), Brazil;
Bulgarian Ministry of Education and Science, within the National Roadmap for Research Infrastructures 2020-2027 (object CERN), Bulgaria;
Ministry of Education of China (MOEC) , Ministry of Science \& Technology of China (MSTC) and National Natural Science Foundation of China (NSFC), China;
Ministry of Science and Education and Croatian Science Foundation, Croatia;
Centro de Aplicaciones Tecnol\'{o}gicas y Desarrollo Nuclear (CEADEN), Cubaenerg\'{\i}a, Cuba;
Ministry of Education, Youth and Sports of the Czech Republic, Czech Republic;
The Danish Council for Independent Research | Natural Sciences, the VILLUM FONDEN and Danish National Research Foundation (DNRF), Denmark;
Helsinki Institute of Physics (HIP), Finland;
Commissariat \`{a} l'Energie Atomique (CEA) and Institut National de Physique Nucl\'{e}aire et de Physique des Particules (IN2P3) and Centre National de la Recherche Scientifique (CNRS), France;
Bundesministerium f\"{u}r Bildung und Forschung (BMBF) and GSI Helmholtzzentrum f\"{u}r Schwerionenforschung GmbH, Germany;
General Secretariat for Research and Technology, Ministry of Education, Research and Religions, Greece;
National Research, Development and Innovation Office, Hungary;
Department of Atomic Energy Government of India (DAE), Department of Science and Technology, Government of India (DST), University Grants Commission, Government of India (UGC) and Council of Scientific and Industrial Research (CSIR), India;
National Research and Innovation Agency - BRIN, Indonesia;
Istituto Nazionale di Fisica Nucleare (INFN), Italy;
Japanese Ministry of Education, Culture, Sports, Science and Technology (MEXT) and Japan Society for the Promotion of Science (JSPS) KAKENHI, Japan;
Consejo Nacional de Ciencia (CONACYT) y Tecnolog\'{i}a, through Fondo de Cooperaci\'{o}n Internacional en Ciencia y Tecnolog\'{i}a (FONCICYT) and Direcci\'{o}n General de Asuntos del Personal Academico (DGAPA), Mexico;
Nederlandse Organisatie voor Wetenschappelijk Onderzoek (NWO), Netherlands;
The Research Council of Norway, Norway;
Pontificia Universidad Cat\'{o}lica del Per\'{u}, Peru;
Ministry of Science and Higher Education, National Science Centre and WUT ID-UB, Poland;
Korea Institute of Science and Technology Information and National Research Foundation of Korea (NRF), Republic of Korea;
Ministry of Education and Scientific Research, Institute of Atomic Physics, Ministry of Research and Innovation and Institute of Atomic Physics and Universitatea Nationala de Stiinta si Tehnologie Politehnica Bucuresti, Romania;
Ministerstvo skolstva, vyskumu, vyvoja a mladeze SR, Slovakia;
National Research Foundation of South Africa, South Africa;
Swedish Research Council (VR) and Knut \& Alice Wallenberg Foundation (KAW), Sweden;
European Organization for Nuclear Research, Switzerland;
Suranaree University of Technology (SUT), National Science and Technology Development Agency (NSTDA) and National Science, Research and Innovation Fund (NSRF via PMU-B B05F650021), Thailand;
Turkish Energy, Nuclear and Mineral Research Agency (TENMAK), Turkey;
National Academy of  Sciences of Ukraine, Ukraine;
Science and Technology Facilities Council (STFC), United Kingdom;
National Science Foundation of the United States of America (NSF) and United States Department of Energy, Office of Nuclear Physics (DOE NP), United States of America.
In addition, individual groups or members have received support from:
Czech Science Foundation (grant no. 23-07499S), Czech Republic;
FORTE project, reg.\ no.\ CZ.02.01.01/00/22\_008/0004632, Czech Republic, co-funded by the European Union, Czech Republic;
European Research Council (grant no. 950692), European Union;
Deutsche Forschungs Gemeinschaft (DFG, German Research Foundation) ``Neutrinos and Dark Matter in Astro- and Particle Physics'' (grant no. SFB 1258), Germany;
ICSC - National Research Center for High Performance Computing, Big Data and Quantum Computing and FAIR - Future Artificial Intelligence Research, funded by the NextGenerationEU program (Italy).

\end{acknowledgement}

\bibliographystyle{utphys}   
\bibliography{bibliography}

\newpage
\appendix

%
%

\section{The ALICE Collaboration}
\label{app:collab}
\begin{flushleft} 
\small

I.J.~Abualrob\,\orcidlink{0009-0005-3519-5631}\,$^{\rm 114}$, 
S.~Acharya\,\orcidlink{0000-0002-9213-5329}\,$^{\rm 50}$, 
G.~Aglieri Rinella\,\orcidlink{0000-0002-9611-3696}\,$^{\rm 32}$, 
L.~Aglietta\,\orcidlink{0009-0003-0763-6802}\,$^{\rm 24}$, 
M.~Agnello\,\orcidlink{0000-0002-0760-5075}\,$^{\rm 29}$, 
N.~Agrawal\,\orcidlink{0000-0003-0348-9836}\,$^{\rm 25}$, 
Z.~Ahammed\,\orcidlink{0000-0001-5241-7412}\,$^{\rm 134}$, 
S.~Ahmad\,\orcidlink{0000-0003-0497-5705}\,$^{\rm 15}$, 
I.~Ahuja\,\orcidlink{0000-0002-4417-1392}\,$^{\rm 36}$, 
Z.~Akbar$^{\rm 81}$
% ...

\section*{Affiliation Notes}

$^{\rm I}$ Deceased\\
$^{\rm II}$ Also at: Max-Planck-Institut fur Physik, Munich, Germany\\
$^{\rm III}$ Also at: Czech Technical University in Prague (CZ)\\
$^{\rm IV}$ Also at: Italian National Agency for New Technologies, Energy and Sustainable Economic Development (ENEA), Bologna, Italy\\
$^{\rm V}$ Also at: Instituto de Fisica da Universidade de Sao Paulo\\
$^{\rm VI}$ Also at: Dipartimento DET del Politecnico di Torino, Turin, Italy\\
$^{\rm VII}$ Also at: Department of Applied Physics, Aligarh Muslim University, Aligarh, India\\
$^{\rm VIII}$ Also at: Institute of Theoretical Physics, University of Wroclaw, Poland\\
$^{\rm IX}$ Also at: Facultad de Ciencias, Universidad Nacional Aut\'{o}noma de M\'{e}xico, Mexico City, Mexico\\

\section*{Collaboration Institutes}

$^{1}$ A.I. Alikhanyan National Science Laboratory (Yerevan Physics Institute) Foundation, Yerevan, Armenia\\
$^{2}$ AGH University of Krakow, Cracow, Poland\\
$^{3}$ Bogolyubov Institute for Theoretical Physics, National Academy of Sciences of Ukraine, Kyiv, Ukraine\\
$^{4}$ Bose Institute, Department of Physics  and Centre for Astroparticle Physics and Space Science (CAPSS), Kolkata, India\\
$^{5}$ California Polytechnic State University, San Luis Obispo, California, United States\\
$^{6}$ Central China Normal University, Wuhan, China\\
$^{7}$ Centro de Aplicaciones Tecnol\'{o}gicas y Desarrollo Nuclear (CEADEN), Havana, Cuba\\
$^{8}$ Centro de Investigaci\'{o}n y de Estudios Avanzados (CINVESTAV), Mexico City and M\'{e}rida, Mexico\\
$^{9}$ Chicago State University, Chicago, Illinois, United States\\
$^{10}$ China Nuclear Data Center, China Institute of Atomic Energy, Beijing, China\\
$^{11}$ China University of Geosciences, Wuhan, China\\
$^{12}$ Chungbuk National University, Cheongju, Republic of Korea\\
$^{13}$ Comenius University Bratislava, Faculty of Mathematics, Physics and Informatics, Bratislava, Slovak Republic\\
$^{14}$ Creighton University, Omaha, Nebraska, United States\\
$^{15}$ Department of Physics, Aligarh Muslim University, Aligarh, India\\
$^{16}$ Department of Physics, Pusan National University, Pusan, Republic of Korea\\
$^{17}$ Department of Physics, Sejong University, Seoul, Republic of Korea\\
$^{18}$ Department of Physics, University of California, Berkeley, California, United States\\
$^{19}$ Department of Physics, University of Oslo, Oslo, Norway\\
$^{20}$ Department of Physics and Technology, University of Bergen, Bergen, Norway\\
$^{21}$ Dipartimento di Fisica, Universit\`{a} di Pavia, Pavia, Italy\\
$^{22}$ Dipartimento di Fisica dell'Universit\`{a} and Sezione INFN, Cagliari, Italy\\
$^{23}$ Dipartimento di Fisica dell'Universit\`{a} and Sezione INFN, Trieste, Italy\\
$^{24}$ Dipartimento di Fisica dell'Universit\`{a} and Sezione INFN, Turin, Italy\\
$^{25}$ Dipartimento di Fisica e Astronomia dell'Universit\`{a} and Sezione INFN, Bologna, Italy\\
$^{26}$ Dipartimento di Fisica e Astronomia dell'Universit\`{a} and Sezione INFN, Catania, Italy\\
$^{27}$ Dipartimento di Fisica e Astronomia dell'Universit\`{a} and Sezione INFN, Padova, Italy\\
$^{28}$ Dipartimento di Fisica `E.R.~Caianiello' dell'Universit\`{a} and Gruppo Collegato INFN, Salerno, Italy\\
$^{29}$ Dipartimento DISAT del Politecnico and Sezione INFN, Turin, Italy\\
$^{30}$ Dipartimento di Scienze MIFT, Universit\`{a} di Messina, Messina, Italy\\
$^{31}$ Dipartimento Interateneo di Fisica `M.~Merlin' and Sezione INFN, Bari, Italy\\
$^{32}$ European Organization for Nuclear Research (CERN), Geneva, Switzerland\\
$^{33}$ Faculty of Electrical Engineering, Mechanical Engineering and Naval Architecture, University of Split, Split, Croatia\\
$^{34}$ Faculty of Nuclear Sciences and Physical Engineering, Czech Technical University in Prague, Prague, Czech Republic\\
$^{35}$ Faculty of Physics, Sofia University, Sofia, Bulgaria\\
$^{36}$ Faculty of Science, P.J.~\v{S}af\'{a}rik University, Ko\v{s}ice, Slovak Republic\\
$^{37}$ Faculty of Technology, Environmental and Social Sciences, Bergen, Norway\\
$^{38}$ Frankfurt Institute for Advanced Studies, Johann Wolfgang Goethe-Universit\"{a}t Frankfurt, Frankfurt, Germany\\
$^{39}$ Fudan University, Shanghai, China\\
$^{40}$ Gangneung-Wonju National University, Gangneung, Republic of Korea\\
$^{41}$ Gauhati University, Department of Physics, Guwahati, India\\
$^{42}$ Helmholtz-Institut f\"{u}r Strahlen- und Kernphysik, Rheinische Friedrich-Wilhelms-Universit\"{a}t Bonn, Bonn, Germany\\
$^{43}$ Helsinki Institute of Physics (HIP), Helsinki, Finland\\
$^{44}$ High Energy Physics Group,  Universidad Aut\'{o}noma de Puebla, Puebla, Mexico\\
$^{45}$ Horia Hulubei National Institute of Physics and Nuclear Engineering, Bucharest, Romania\\
$^{46}$ HUN-REN Wigner Research Centre for Physics, Budapest, Hungary\\
$^{47}$ Indian Institute of Technology Bombay (IIT), Mumbai, India\\
$^{48}$ Indian Institute of Technology Indore, Indore, India\\
$^{49}$ INFN, Laboratori Nazionali di Frascati, Frascati, Italy\\
$^{50}$ INFN, Sezione di Bari, Bari, Italy\\
$^{51}$ INFN, Sezione di Bologna, Bologna, Italy\\
$^{52}$ INFN, Sezione di Cagliari, Cagliari, Italy\\
$^{53}$ INFN, Sezione di Catania, Catania, Italy\\
$^{54}$ INFN, Sezione di Padova, Padova, Italy\\
$^{55}$ INFN, Sezione di Pavia, Pavia, Italy\\
$^{56}$ INFN, Sezione di Torino, Turin, Italy\\
$^{57}$ INFN, Sezione di Trieste, Trieste, Italy\\
$^{58}$ Inha University, Incheon, Republic of Korea\\
$^{59}$ Institute for Gravitational and Subatomic Physics (GRASP), Utrecht University/Nikhef, Utrecht, Netherlands\\
$^{60}$ Institute of Experimental Physics, Slovak Academy of Sciences, Ko\v{s}ice, Slovak Republic\\
$^{61}$ Institute of Physics, Homi Bhabha National Institute, Bhubaneswar, India\\
$^{62}$ Institute of Physics of the Czech Academy of Sciences, Prague, Czech Republic\\
$^{63}$ Institute of Space Science (ISS), Bucharest, Romania\\
$^{64}$ Institut f\"{u}r Kernphysik, Johann Wolfgang Goethe-Universit\"{a}t Frankfurt, Frankfurt, Germany\\
$^{65}$ Instituto de Ciencias Nucleares, Universidad Nacional Aut\'{o}noma de M\'{e}xico, Mexico City, Mexico\\
$^{66}$ Instituto de F\'{i}sica, Universidade Federal do Rio Grande do Sul (UFRGS), Porto Alegre, Brazil\\
$^{67}$ Instituto de F\'{\i}sica, Universidad Nacional Aut\'{o}noma de M\'{e}xico, Mexico City, Mexico\\
$^{68}$ iThemba LABS, National Research Foundation, Somerset West, South Africa\\
$^{69}$ Jeonbuk National University, Jeonju, Republic of Korea\\
$^{70}$ Johann-Wolfgang-Goethe Universit\"{a}t Frankfurt Institut f\"{u}r Informatik, Fachbereich Informatik und Mathematik, Frankfurt, Germany\\
$^{71}$ Korea Institute of Science and Technology Information, Daejeon, Republic of Korea\\
$^{72}$ Laboratoire de Physique Subatomique et de Cosmologie, Universit\'{e} Grenoble-Alpes, CNRS-IN2P3, Grenoble, France\\
$^{73}$ Lawrence Berkeley National Laboratory, Berkeley, California, United States\\
$^{74}$ Lund University Department of Physics, Division of Particle Physics, Lund, Sweden\\
$^{75}$ Nagasaki Institute of Applied Science, Nagasaki, Japan\\
$^{76}$ Nara Women{'}s University (NWU), Nara, Japan\\
$^{77}$ National and Kapodistrian University of Athens, School of Science, Department of Physics , Athens, Greece\\
$^{78}$ National Centre for Nuclear Research, Warsaw, Poland\\
$^{79}$ National Institute of Science Education and Research, Homi Bhabha National Institute, Jatni, India\\
$^{80}$ National Nuclear Research Center, Baku, Azerbaijan\\
$^{81}$ National Research and Innovation Agency - BRIN, Jakarta, Indonesia\\
$^{82}$ Niels Bohr Institute, University of Copenhagen, Copenhagen, Denmark\\
$^{83}$ Nikhef, National institute for subatomic physics, Amsterdam, Netherlands\\
$^{84}$ Nuclear Physics Group, STFC Daresbury Laboratory, Daresbury, United Kingdom\\
$^{85}$ Nuclear Physics Institute of the Czech Academy of Sciences, Husinec-\v{R}e\v{z}, Czech Republic\\
$^{86}$ Oak Ridge National Laboratory, Oak Ridge, Tennessee, United States\\
$^{87}$ Ohio State University, Columbus, Ohio, United States\\
$^{88}$ Physics department, Faculty of science, University of Zagreb, Zagreb, Croatia\\
$^{89}$ Physics Department, Panjab University, Chandigarh, India\\
$^{90}$ Physics Department, University of Jammu, Jammu, India\\
$^{91}$ Physics Program and International Institute for Sustainability with Knotted Chiral Meta Matter (WPI-SKCM$^{2}$), Hiroshima University, Hiroshima, Japan\\
$^{92}$ Physikalisches Institut, Eberhard-Karls-Universit\"{a}t T\"{u}bingen, T\"{u}bingen, Germany\\
$^{93}$ Physikalisches Institut, Ruprecht-Karls-Universit\"{a}t Heidelberg, Heidelberg, Germany\\
$^{94}$ Physik Department, Technische Universit\"{a}t M\"{u}nchen, Munich, Germany\\
$^{95}$ Politecnico di Bari and Sezione INFN, Bari, Italy\\
$^{96}$ Research Division and ExtreMe Matter Institute EMMI, GSI Helmholtzzentrum f\"ur Schwerionenforschung GmbH, Darmstadt, Germany\\
$^{97}$ Saga University, Saga, Japan\\
$^{98}$ Saha Institute of Nuclear Physics, Homi Bhabha National Institute, Kolkata, India\\
$^{99}$ School of Physics and Astronomy, University of Birmingham, Birmingham, United Kingdom\\
$^{100}$ Secci\'{o}n F\'{\i}sica, Departamento de Ciencias, Pontificia Universidad Cat\'{o}lica del Per\'{u}, Lima, Peru\\
$^{101}$ Stefan Meyer Institut f\"{u}r Subatomare Physik (SMI), Vienna, Austria\\
$^{102}$ SUBATECH, IMT Atlantique, Nantes Universit\'{e}, CNRS-IN2P3, Nantes, France\\
$^{103}$ Sungkyunkwan University, Suwon City, Republic of Korea\\
$^{104}$ Suranaree University of Technology, Nakhon Ratchasima, Thailand\\
$^{105}$ Technical University of Ko\v{s}ice, Ko\v{s}ice, Slovak Republic\\
$^{106}$ The Henryk Niewodniczanski Institute of Nuclear Physics, Polish Academy of Sciences, Cracow, Poland\\
$^{107}$ The University of Texas at Austin, Austin, Texas, United States\\
$^{108}$ Universidad Aut\'{o}noma de Sinaloa, Culiac\'{a}n, Mexico\\
$^{109}$ Universidade de S\~{a}o Paulo (USP), S\~{a}o Paulo, Brazil\\
$^{110}$ Universidade Estadual de Campinas (UNICAMP), Campinas, Brazil\\
$^{111}$ Universidade Federal do ABC, Santo Andre, Brazil\\
$^{112}$ Universitatea Nationala de Stiinta si Tehnologie Politehnica Bucuresti, Bucharest, Romania\\
$^{113}$ University of Derby, Derby, United Kingdom\\
$^{114}$ University of Houston, Houston, Texas, United States\\
$^{115}$ University of Jyv\"{a}skyl\"{a}, Jyv\"{a}skyl\"{a}, Finland\\
$^{116}$ University of Kansas, Lawrence, Kansas, United States\\
$^{117}$ University of Liverpool, Liverpool, United Kingdom\\
$^{118}$ University of Science and Technology of China, Hefei, China\\
$^{119}$ University of Silesia in Katowice, Katowice, Poland\\
$^{120}$ University of South-Eastern Norway, Kongsberg, Norway\\
$^{121}$ University of Tennessee, Knoxville, Tennessee, United States\\
$^{122}$ University of the Witwatersrand, Johannesburg, South Africa\\
$^{123}$ University of Tokyo, Tokyo, Japan\\
$^{124}$ University of Tsukuba, Tsukuba, Japan\\
$^{125}$ Universit\"{a}t M\"{u}nster, Institut f\"{u}r Kernphysik, M\"{u}nster, Germany\\
$^{126}$ Universit\'{e} Clermont Auvergne, CNRS/IN2P3, LPC, Clermont-Ferrand, France\\
$^{127}$ Universit\'{e} de Lyon, CNRS/IN2P3, Institut de Physique des 2 Infinis de Lyon, Lyon, France\\
$^{128}$ Universit\'{e} de Strasbourg, CNRS, IPHC UMR 7178, F-67000 Strasbourg, France, Strasbourg, France\\
$^{129}$ Universit\'{e} Paris-Saclay, Centre d'Etudes de Saclay (CEA), IRFU, D\'{e}partment de Physique Nucl\'{e}aire (DPhN), Saclay, France\\
$^{130}$ Universit\'{e}  Paris-Saclay, CNRS/IN2P3, IJCLab, Orsay, France\\
$^{131}$ Universit\`{a} degli Studi di Foggia, Foggia, Italy\\
$^{132}$ Universit\`{a} del Piemonte Orientale, Vercelli, Italy\\
$^{133}$ Universit\`{a} di Brescia, Brescia, Italy\\
$^{134}$ Variable Energy Cyclotron Centre, Homi Bhabha National Institute, Kolkata, India\\
$^{135}$ Warsaw University of Technology, Warsaw, Poland\\
$^{136}$ Wayne State University, Detroit, Michigan, United States\\
$^{137}$ Yale University, New Haven, Connecticut, United States\\
$^{138}$ Yildiz Technical University, Istanbul, Turkey\\
$^{139}$ Yonsei University, Seoul, Republic of Korea\\
$^{140}$ Affiliated with an institute formerly covered by a cooperation agreement with CERN\\
$^{141}$ Affiliated with an international laboratory covered by a cooperation agreement with CERN.\\

\end{flushleft} 

\end{document}